\documentclass[aps,prl,floatfix,amsmath,amssymb,groupedaddress,showpacs,twocolumn]{revtex4}
\usepackage[dvips]{graphicx}

\newcommand{\Act}{\mathcal{S}}
\newcommand{\Ham}{\mathcal{H}}
\newcommand{\Order}{\mathcal{O}}

\newcommand{\be}{\begin{equation}}
\newcommand{\ee}{\end{equation}}
\newcommand{\punc}[1]{\;\mathrm{#1}}

\newcommand{\e}{\mathrm{e}}
\newcommand{\ii}{\mathrm{i}}
\newcommand{\dd}{\,\mathrm{d}}

\newcommand{\refeq}[1]{Eq.~(\ref{#1})}

\newcommand{\phc}{+\mathrm{h.c.}}

\newcommand{\Lv}{\mathbf{L}}
\newcommand{\kv}{\mathbf{k}}
\newcommand{\xv}{\mathbf{x}}

\newcommand{\nh}{\hat{n}}

\newcommand{\bra}[1]{\langle #1|}
\newcommand{\ket}[1]{|#1\rangle}

\newcommand{\spaceitimeint}{\int d^d\xv\, d\tau\:}

\newcommand{\sub}[1]{_{\mathrm{#1}}}

\begin{document}
\setlength{\unitlength}{1mm}

\title{Excited state spectra at the superfluid-insulator
transition\\ out of paired condensates}

\author{Stephen Powell}
\affiliation{Department of Physics, Yale University, New Haven, CT
06520-8120}

\author{Subir Sachdev}
\affiliation{Department of Physics, Harvard University, Cambridge MA
02138}

\begin{abstract}
We describe gapped single-particle and collective excitations across
a superfluid to insulator quantum phase transition of particles
(bosons or fermions) in a periodic potential, with an even number of
particles per unit cell. We demonstrate that the dynamics is
controlled by a {\em quantum impurity\/} problem of a localized
particle interacting with the bulk critical modes. Critical
exponents are determined by a renormalization group analysis. We
discuss applications to spin oscillations of ultracold atoms in
optical lattices, and to the electronic phases in the cuprate and
related compounds.
\end{abstract}





\date{\today}

\maketitle

A number of experiments have observed coherent spin oscillations of
trapped, ultracold spinor bosonic atoms. With a superfluid ground
state \cite{
ketterle1,chapman1,kurn,chapman2,sengstock1,sengstock2}, these
oscillations are well described by the classical (Gross-Pitaevski)
equations which control the time evolution of the multi-component
atomic condensate. Recent experiments \cite{bloch1,bloch2} have also
observed oscillations in a Mott insulating state obtained by placing
the atoms in an optical lattice; in this situation, the spin
oscillations can be viewed as the Rabi precession
\cite{bloch1,zhang} between sharp quantum states localized within
each minimum of the optical lattice.

In this paper, we investigate the connection between these two
disparate pictures of the spin oscillations, by describing ground
state spin correlations at the quantum critical point separating the
superfluid and insulating phases. Bosonic atoms with total spin
$F=1$ and $F=2$ display a remarkably rich variety of superfluid and
insulating phases \cite{demler1,kimura,fazio,lewenstein,ryan}, and
have a corresponding plethora of universality classes of quantum
phase transitions separating such phases. Superfluid-insulator
transitions are also possible for paired fermionic atoms in an
optical lattice. We defer a more complete classification of the spin
dynamics at such transitions to a forthcoming paper. Here we focus
on a class, with an even integer number of particles per unit cell,
which displays non-trivial collective behavior induced by a strong
coupling between the spin excitations and the critical number and
phase fluctuations of the superfluid-insulator transition; the class
includes both the fermionic and bosonic cases. The problem is mapped
exactly onto a quantum `impurity' problem, which couples a single
localized spin excitation (the `impurity') to the bulk critical
modes; a solution using the renormalization group yields new
critical exponents and scaling functions. Our theory has some
analogies to simpler models of the Kondo and X-ray edge effect in
metals, and opens the way towards observing strongly-coupled quantum
impurity physics in ultracold atom systems.

Our results also apply to the superfluid-insulator transition in
electronic systems, with an even integer number of electrons per
unit cell. Such a situation can arise in the cuprate or related
compounds, with a periodic potential generated spontaneously by
`stripe/checkerboard' or charge density wave order: recent
experiments in insulating spin ladder compounds \cite{abbamonte}
have shown that each unit cell contains a pair of holes. Our results
predict the frequency dependence of the electron photoemission
spectrum across a superfluid-insulator transition in which the
`stripe' order is present on both sides of the transition; in other
words, for a transition between a modulated insulator and a
supersolid. The predictions of the spectrum are for gapped single
particle excitations at positions of the gap maxima or minima in the
Brillouin zone {\em e.g.\/} at the analog of the `antinodal' points.

We will study superfluid-insulator transitions at which the energy
gap to both single-particle and spin excitations remains nonzero at
the transition. The superfluid order parameter for the transition is
then necessarily a spin singlet. The order parameter also carries a
nonzero particle number, or `charge' $Q$. Here, and henceforth,
`particle' refers to either a single ultracold bosonic or fermionic
atom, or an electron (but {\em not\/} a Cooper pair). For $F \neq 0$
particles, this means that the simplest case has an order parameter,
$\Psi$, with $Q=2$ and $F=0$, corresponding to the annihilation
operator for a spin-singlet pair of particles, {\em e.g.\/} a Cooper
pair.

For definiteness, we will develop our results in the context of a
Bose-Hubbard model for $F \neq 0$ bosons in an optical
lattice, and indicate the generalization to the fermionic case later.
For this model, we consider the transition
from a `spin-singlet insulator' (SSI), a Mott insulator with an even
number of atoms per lattice site and no spin order, to a
`spin-singlet condensate' (SSC), in which singlet pairs of bosons
have condensed, but there is no single boson condensate. We will
begin with a simple mean-field theory of the Bose-Hubbard model, and
then turn to a field theory of the critical properties of this and
the corresponding fermionic model.

The lattice bosons are annihilated with operators $a_{i,m}$ on
lattice site $i$ and spin projection $m=-F \ldots F$. The
Bose-Hubbard Hamiltonian can then be written as $\Ham = - t T + V$,
where $T$ is the kinetic energy term,
\be
T = \sum_{\langle
i,j\rangle,m} (a_{i,m}^\dag a_{j,m} \phc) \punc{,}
\ee
and $V$ is the on-site interaction:
\be
\label{OnsiteV}
V = \sum_i
\left[U(\nh_i-N)^2 + J |\Lv_i|^2\right]\punc{,}
\ee
where $\nh_i = \sum_m a^\dagger_m a_m$ is the boson number operator on
site $i$, and $\Lv_i$ is the total spin operator on site $i$.
We have made the
spin-independent part of the interaction explicitly symmetric around
$N$ particles per site. For $F=1 $, the final term is the most
general spin-dependent interaction, but further terms are necessary
for higher spin. To favor spin-singlet pairing in the ground state,
we require $N$ to be even, and $J > 0$.

In the case when $t = 0$, the Hamiltonian is simply a sum of
terms acting on a single site, containing only the commuting
operators $\nh$ and $|\Lv|^2$. The ground state on each site
is therefore a spin singlet of $N$ bosons.

An appropriate mean-field Hamiltonian is $\Ham\sub{mf} = V -
T_\psi - T_\Psi - T_\Phi$, where $V$ is the same on-site
interaction as in \refeq{OnsiteV}. $T_\psi$ is the standard
mean-field decoupling of the hopping term, generalized to
the case with spin,
\be
T_\psi = \sum_i \left[ \psi_m a_{i,m}^\dag + \psi_m^*
a_{i,m}\right]\punc{,}
\ee
where $\psi_m$ is a (c-number) constant vector, which will
be used as a variational parameter. The remaining terms
allow for the possibility of a spin-singlet condensate
through the parameters $\Psi$ and $\Phi$:
\be
\label{TPsi} T_\Psi = \sum_{i,m} (-1)^{F+m} \left[ \Psi a_{i,m}^\dag
a_{i,-m}^\dag + \Psi^* a_{i,m}a_{i,-m}\right]\punc{,}
\ee
and
\be
T_\Phi = \sum_{\langle i,j\rangle,m} (-1)^{F+m} \left[ \Phi a_{i,m}^
\dag a_{j,-m}^\dag + \Phi^* a_{i,m}a_{j,-m}\right]\punc{,}
\ee
where the factors of $(-1)^{F+m} $ are Clebsch-Gordan
coefficients that cause the boson operators to form spin-
singlet pairs.

\newcommand{\mf}{\mathrm{mf}} \newcommand{\up}[1]{^{(#1)}}
We now use the ground state of $\Ham\sub{mf}$, which we
denote $\ket{\mf}$, as a variational ansatz and define
\be
E\sub{mf}(\psi_m, \Psi, \Phi) =
\bra{\mf}\Ham\ket{\mf}\punc{,}
\ee
which should be minimized by varying the three parameters.
If this minimum occurs for vanishing values of all three
parameters, then $\ket\mf$ breaks no symmetries and the Mott
insulator is favored. A nonzero value for $\psi_m$ at the
minimum corresponds to a simple `polar condensate' (PC),
breaking spin-rotation symmetry; vanishing $\psi_m$
but nonzero values of $\Psi$ and/or $\Phi$ corresponds to a
paired spin-singlet condensate.

\begin{figure}
\includegraphics{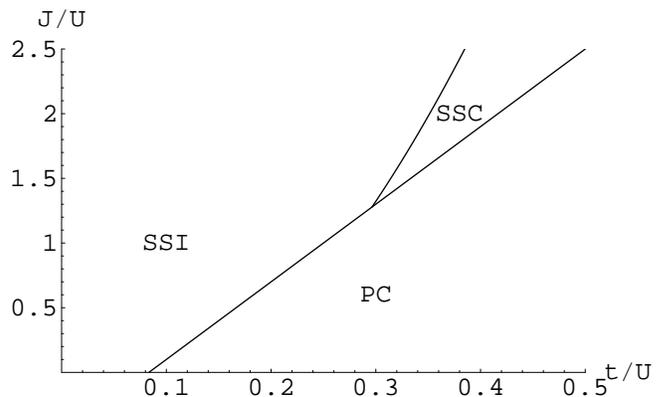}
\caption{\label{MeanFieldPhaseDiagram}Phase diagram
calculated using mean-field theory. The calculation has been
performed for spin $F=1$ and filling factor $N=2$. The
horizontal and vertical axes give the tunneling strength $t$
and the spin-dependent interaction $J$, both in units of the
spin-independent part of the interaction, $U$.}
\end{figure}

Since $\Ham\sub{mf}$ contains terms (within $T_\Phi$) that
link adjacent sites, it cannot be straightforwardly
diagonalized, as in the standard mean-field theory for the
spinless Bose-Hubbard model. To find the phase boundaries,
however, we need only terms up to quadratic order in the
variational parameters, which can be found using
perturbation theory. Figure \ref
{MeanFieldPhaseDiagram} shows the phase
boundaries so obtained.

We now turn to an analysis of the correlation functions of
gapped spin-carrying modes across the SSI/SSC transition. We
will assume we are in a regime in which the lowest
excitation with a nonzero spin is created by an elementary
boson operator, $\psi_m$, with $Q=1$ and $F\neq 0$, with
$m=-F, \ldots, F$.

General symmetry arguments can be used to deduce the low-energy theory coupling $\Psi$ to $\psi_m$ in the
vicinity of the superfluid-insulator transition. For the $
\Psi$ field alone, we have the familiar $|\Psi |^4$ field
theory of the superfluid-insulator transition,
\begin{equation}
\mathcal{S}_\Psi = \spaceitimeint \left(|\partial \Psi|^2 +
r_\Psi | \Psi|^2 + \frac{u_\Psi}{2}|\Psi|^4 +
\,\cdots\right)
\end{equation}
in $d$ spatial dimensions with co-ordinate $\xv$, imaginary time
$\tau$, and $(d+1)$-dimensional derivative $\partial^2 =
\partial_\tau^ 2 + \nabla^2$. The even
integer number of bosons per lattice site ensures that there is
particle-hole symmetry in $\mathcal{S}_\Psi$ in the vicinity of the
critical point.

The same particle-hole symmetry applies to the gapped $
\psi_m$ field, but it is convenient to treat separately the
particle-like (with $Q=1$) and hole-like (with $Q=-1$)
excitations. The long-time behavior will be governed by
those excitations just above the gap $\lambda$, for which
the `nonrelativistic' limit can be taken and the action
written
\begin{multline}
\Act_{\psi} = \sum_m \spaceitimeint \bigg[ p^{\dagger}_m \left(
\ii \partial_t - \frac{1}{2m_p} \nabla^2 + \lambda\right) p_m \\
+ h^{\dagger}_m \left(\ii\partial_t - \frac{1}{2m_h} \nabla^2+
\lambda\right) h_m\bigg]\punc{,}
\end{multline}
where $p_m$ and $h_m$ are fields describing particle and hole
excitations of $\psi_m \sim u p_m + v (-1)^{F+m} h^{\dagger}_{-m}$,
where $u$ and $v$ are coefficients chosen to that there is no
$p_m h_m$ term in $\Act_{\psi}$. The even integer density constraint
requires that the gap $\lambda$ be the same in the particle and hole
sectors, but the masses $m_{p,h}$ are, in general, allowed to be
different.

The coupling between $\Psi$ and $\psi_m$ allows the conversion of a
particle to a hole along with the creation of a pair:
\be
\Act_g = g
\sum_m \spaceitimeint \left( \Psi^\dagger h^{\dagger}_{m} p_m +
p^{\dagger}_m h_{m} \Psi \right)\punc{.}
\ee
Our central results for the gapped excitations at the
superfluid-insulator transition follow from the quantum field theory
defined by $\Act_\Psi + \Act_\psi +\Act_g$, and we will describe its
properties below.

However, before we do so, we note that the same theory applies also
to the superfluid-insulator transitions of fermions, with an even
integer number of {\em fermions\/} per site: the only change is that
the single particle excitations $p_m$, $h_m$ are fermionic. For a
derivation for $F=1/2$ electrons, note that the filling $N=2$ per
site requires the underlying model to have at least two bands to
prevent a trivial filled-band ground state. Suitable orthogonal
linear combinations of the electron annihilation operators in these
bands yield the $p_m$ and $\varepsilon_{mm'} h^\dagger_{m'}$ respectively.
($\varepsilon$ is the $2 \times 2$ antisymmetric tensor).

We are interested here in the spectral functions of gapped
excitations created by exciting a finite number of $p$ and $h$
quanta: at $T=0$, these will have thresholds in the spectral density
at integer multiples of $\lambda$. As in previous work on spin
ordering transitions \cite{stv}, we now demonstrate that the
structure of these threshold singularities reduces to the solution
of an associated quantum impurity problem.

$\Act_\Psi$ is isotropic in $d+1$ space- time dimensions, and so, at
its critical point, is invariant under scaling transformations with
dynamic exponent $z=1$. When such a rescaling transformation is
applied to $\Act_\psi$, it is clear that the coefficients of the
dispersion $(2m_{p,h})^{-1}$ have scaling dimensions $-1$, and are
therefore irrelevant. The critical behavior is therefore given by
$(2m_{p,h})^{-1} = 0$, which describes a static impurity coupling to
the bulk theory of $\Act_\Psi$.

Applying the same scaling analysis to the coupling in $
\Act_g$ shows that the scaling dimension of $g$ is
$(3-d)/2$, so that it is relevant for $d < 3$. Below we will
address the case of two (spatial) dimensions, and describe
an expansion in $\epsilon = 3-d$ for the critical exponents.

We concentrate first on single-particle excitations with charge $Q =
1$ and spin $F$. On the insulating side of the transition, where
$r_\Psi > 0$, the Green function for the $\psi_m$ field, $G^\psi$,
will have a pole corresponding to the stable excitation at frequency
$\omega = \lambda$, and a continuum for $\omega > \lambda +
\sqrt{r_\Psi}$, at frequencies large enough for a particle to
produce a hole along with a pair. As the transition is approached,
$r_\Psi$ will become smaller while $\lambda$ remains fixed, and the
continuum will start closer to the pole. On the superfluid side, the
action described in $ \Act_\Psi$ must be rewritten in terms of
amplitude and phase modes of the condensate. Away from the
transition, the amplitude mode remains gapped, while the phase mode
is the gapless Goldstone mode corresponding to broken phase-
rotation symmetry in the superfluid. Coupling between this gapless
mode and $ \psi_m$ will cause the continuum in $G^ \psi$ to appear
for frequencies just above $\lambda$. The quasiparticle pole
remains, however, due to the factors of momentum appearing in the
matrix element for coupling to the Goldstone mode, which cause the
scattering rate to decrease as some power of $\omega - \lambda$.

Exactly at the transition, by contrast, the coupling is not
restricted by the Goldstone theorem and the critical $\Psi$
excitations qualitatively modify the structure of $G^\psi$.
The same is true of higher-order correlation functions, such
as that for $p_m h_{m'}$, which is the simplest
spin-carrying but charge-neutral combination.

These correlation functions can be found
using a renormalization group (RG) calculation based on the
formulation as a quantum impurity problem. Our approach is
to perform a rescaling, which leaves $\Act_\Psi$ invariant,
to relate the correlation function evaluated at a frequency
just above the gap, say $\omega = \lambda + \delta \omega$,
to another frequency, $\omega' = \lambda + \e^{-\ell} \delta
\omega$, where $\ell$ is infinitesimal. The form of the
coupling $\Act_g$ allows the scale-invariance of $\Act_\Psi$
to be used to relate correlators at these two frequencies.

As in the standard RG, the second stage of the calculation
involves restoring the momentum cutoff to its original
value. To lowest order in the coupling $g$ (or, as will subsequently be
shown to be equivalent, in an expansion in $ \epsilon = 3 - d$), the
only diagram that must be calculated for the renormalization of
$G^\psi$ is the self-energy diagram
\be
\Sigma^\psi_1(\ii\omega)
=\:
\parbox{25mm}{
\begin{picture}(25,10)
\put(0,0){\includegraphics{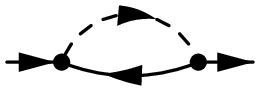}}
\end{picture}
}
\label{SelfEnergyDiagram1}
\ee
where the solid lines represent $\psi_m$ and the dashed line $\Psi$. The constant
part of the diagram $\Sigma^\psi_1(\ii \omega)$, along with a
lower-order `tadpole' diagram that we have omitted, produces a
renormalization of the gap $\lambda $, but this is of no interest
here.

It happens that there are no diagrams renormalizing the coupling $g$
at this order, although such diagrams do appear at higher order, and
can be computed as in Ref.~\cite{japan}.

Since $\psi_m$ has only gapped excitations, it has no effect on the
bulk scaling, which is therefore given by the standard results for a
complex $|\Psi|^4$ theory. Rescaling by the standard phase space
factor $\hat{u} = u_\Psi/S_{d+1}$ (where $S_d = 2/(\Gamma(d/2) (4
\pi)^{d/2})$), there is a fixed point with coupling $\hat{u} =
\epsilon/5 + \Order(\epsilon^2) $ in $3 - \epsilon$ dimensions, at
which the scaling dimension of the field is given by $[\Psi] = 1 -
\frac{\epsilon}{2} + \epsilon^2 /100 + \Order(\epsilon^3)$.

These results for the bulk, along with the self-energy
diagram above, when extended to two loops,
lead to the RG flow equation for the coupling
to the `impurity':
\be
\label{betag}
\frac{\dd \hat{g}}{\dd \ell} = \frac{\epsilon}{2}\hat{g} -
\hat{g}^3 + 2\hat{g}^5 - \frac{\hat{u}^2 \hat{g}}{4}
- \frac{2 \pi^2 \hat{u} \hat{g}^3}{3} + \Order(\hat{g}(\hat{u}, \hat{g}^2)^3)\punc{,}
\ee
where we have defined $\hat{g} =
g (4 \pi)^{(d+1)/2} /\Gamma((d-1)/2)$.
The coupling therefore approaches a fixed-point
value with $\hat{g}^2 = \epsilon/2 - (\pi^2/15 - 49/100) \epsilon^2
+ \mathcal{O}(\epsilon^3 )$, so that the perturbative
expansion at this point is indeed equivalent to an expansion
in $\epsilon$.

Finally, using the wavefunction renormalization of $\psi_m$,
we arrive at the rescaling of the Green function $G^\psi$,
which obeys
\be
G^\psi(\lambda + \delta\omega) = \e^{-y \ell} G^\psi(\lambda
+ \e^{- \ell}\delta\omega)\punc{,}
\ee
where $y = 1 - \hat{g}^2 + \hat{g}^4 + \mathcal{O}(\hat{g}^6)$,
with $\hat{g}$ set to the fixed point of Eq.~(\ref{betag}); this yields
$y=1- \epsilon/2 + (6/25 - \pi^2 /15) \epsilon^2 + O(\epsilon^3)$. This can be
iterated to give $G^\psi(\lambda + \delta\omega) \sim \delta
\omega^{-y} $. Note that, at least to this order, $y < 1$,
so that the quasiparticle pole at $\omega = \lambda$ is
replaced by a power-law threshold singularity. This exponent, determining the
spectral density of the `photoemission' of a single hole or particle at a
band minimum or maximum,
is one of our central results.

A similar calculation applies to the two-particle threshhold singularity
at $\omega = 2 \lambda$. This is associated with
the renormalization of the $T_{mm'} p_m
h_{m'}$ correlation function, where $T_{mm'}$ is an arbitrary matrix.
It is then also necessary to
calculate the following insertion diagram (at one loop order):
\be H_1
=\:
\parbox{20mm}{
\begin{picture}(20,20)
\put(0,0){\includegraphics{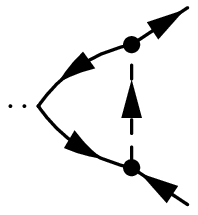}}
\end{picture}
}
\label{InsertionDiagram1}
\ee
which accounts for
the renormalization associated with bringing the particle and hole
operators to the same space-time point.
The resulting calculation is very similar and leads to the
result for the two-particle propagator
\be
\Pi(\omega, \kv=0) \sim (2\lambda - \omega)^{-y_2}\punc{,}
\ee
where the exponent $y_2$ depends upon whether the matrix $T_{mm'}$ is
symmetric or antisymmetric. For bosons, a symmetric
$T$ has $y_2 = 1 - 2 \hat{g}^2 + \mathcal{O} (\hat{g}^6 )$, while
an antisymmetric $T$ has $y_2 = 1$ exactly; these exponents therefore determine
the threshold singularities for excitations
with even and odd total spin $F$, respectively. For fermions, the same
results apply, but with the roles
of symmetric/antisymmetric $T$ reversed.

This paper has presented results for a variety of threshold singularities
in the spectral functions of a system undergoing a superfluid-insulator
transition, with an even number of particles per unit cell. This has direct
application to experiments on ultracold atoms and to the cuprate compounds.
It would be interesting to extend these methods to study non-equilibrium phenomena,
such as those measured in Ref.~\cite{bloch1}, by methods explored in recent
studies of non-equilibrium quantum criticality \cite{powell1,cardy} (which
have so far been limited to 1+1 dimensions).

We thank E.~Demler and R.~Shankar for useful discussions.
This research was supported by the NSF
grants DMR-0537077, DMR-0342157, and DMR-0354517.


\begin{thebibliography}{99}

\bibitem{ketterle1} J.~Stenger {\em et al.},
Nature {\bf 396}, 345 (1998).

\bibitem{chapman1} M.~S.~Chang {\em et al},
Phys. Rev. Lett. {\bf 92}, 140403 (2004).

\bibitem{kurn} J.~M.~Higbie {\em et al},
Phys. Rev. Lett. {\bf 95}, 050401 (2005).

\bibitem{chapman2} M.~S.~Chang {\em et al},
Nature Phys. {\bf 1}, 111 (2005).

\bibitem{sengstock1} J.~Mur-Petit {\em et al},
Phys. Rev. A {\bf 73}, 013629 (2006).

\bibitem{sengstock2} J.~Kronj\"ager {\em et al},
cond-mat/ 0509083.

\bibitem{bloch1} A.~Widera {\em et al},
Phys. Rev. Lett. {\bf 95}, 190405 (2005)

\bibitem{bloch2} F.~Gerbier {\em et al},
Phys. Rev. A {\bf 73}, 041602 (2006).

\bibitem{zhang} H.-J.~Huang and G.-M.~Zhang, cond-mat/0601188.

\bibitem{demler1} A.~Imambekov, M.~Lukin, and E.~Demler, Phys. Rev. A {\bf 68}, 063602 (2003).

\bibitem{kimura} S.~Tsuchiya, S.~Kurihara, and T.~Kimura, Phys. Rev. A {\bf 70}, 043628 (2004).

\bibitem{fazio}  D.~Rossini {\em et al},
J. Phys. B: At. Mol. Opt. Phys. {\bf 39}, S163 (2006).

\bibitem{lewenstein} L.~Zawitkowski {\em et al},
cond-mat/0603273.

\bibitem{ryan} R.~Barnett, A.~Turner, and E.~Demler, cond-mat/ 0607253.

\bibitem{abbamonte} A.~Rusydi {\em et al}, cond-mat/0604101.

\bibitem{stv} S.~Sachdev, M.~Troyer, and M.~Vojta, Phys. Rev. Lett. { \bf 86}, 2617 (2001).

\bibitem{japan} S. Sachdev, Physica C {\bf 357}, 78 (2001).

\bibitem{powell1} K.~Sengupta, S.~Powell, and S.~Sachdev, Phys. Rev. A {\bf 69}, 053616 (2004).

\bibitem{cardy} P.~Calabrese and J.~Cardy, Phys. Rev. Lett. {\bf 96}, 136801 (2006).

\end{thebibliography}
\end{document}